# The reliability of a nutritional meta-analysis study

Karl E. Peace[1], PhD; JingJing Yin[1], PhD; Haresh Rochani[1*], DrPH; Sarbesh Pandeya[1], MPH, S. Stanley Young[2], PhD

[1] Department of Biostatistics, Jiann-Ping Hsu College of Public Health, Statesboro, GA, 30460

[2] CEO, CGSTAT, 3401 Caldwell Drive, Raleigh, NC 27607

Funding: - None

[*]To whom correspondence should be addressed. 501 Forest drive, P.O Box 8015. E-mail: hrochani@georgiasouthern.edu

**Abstract**

**Background**: Many researchers have studied the relationship between diet and health. Specifically, there are papers showing an association between the consumption of sugar-sweetened beverages and Type 2 diabetes. Many meta-analyses use individual (primary) studies that don't adjust for multiple testing or multiple modeling – and thus provide biased estimates of effect. Hence the claims reported in a meta-analysis paper may be unreliable if the primary papers do not ensure unbiased estimates of effect.

**Objective**: Determine (i) the statistical reliability of 10 papers **and** (ii) indirectly the reliability of the meta-analysis study.

**Method**: We obtained copies of each of the 10 primary papers used in a meta-analysis paper and counted the numbers of outcomes, predictors, and covariates. We estimated the size of the potential analysis search space available to the authors of these papers; i.e. the number of comparisons and models available. The potential analysis search space is the number of outcomes times the number of predictors times $2^c$, where c is the number of covariates. Since we noticed that there were differences between predictors and covariates cited in the abstract and in the text, we applied this formula to information found in the abstracts, Space A, as well as the text, Space T, of each primary paper.

**Results**: The median and range of the number of comparisons possible across the primary papers are 6.5 and (2 - 12,288), respectively for the abstracts, and 196,608 and (3,072 -117,117,952), respectively for the texts. Note that the median of 6.5 for Space A is misleading as each primary study has 60-165 foods not mentioned in the abstract.

**Conclusion**: Given that testing is at the 5% level and the number of comparisons is very large, nominal statistical significance is very weak support for a claim. The claims in these papers are not statistically supported and hence are unreliable. Thus, the claims of the meta-analysis paper lack evidentiary confirmation.

**1.0** Summary of logic of this paper

Often a paper is presented as a story where certain items are highlighted; items that do not fit the story are sometimes omitted; the reader must wait to the end to see how it all turns out. Here the items are too complex for a simple story so we outline the internal logic of our study to assist the reader. The paper fills in details and offers justifications.

First, claims coming from observational studies most often do not replicate. Second, using simple counting, we determine search spaces, the number of questions at issue, in the ten primary papers used in a meta-analysis study. Multiple testing and multiple modeling, MTMM, are well-known problems of observational studies that can produce small (invalid) p-values. Although MTMM are well-understood, in practice they are often ignored. Next, we note that if a false positive p-value is used to claim an effect, it derives from a biased estimate of a treatment effect. Finally, a fundamental requirement of a meta-analysis is unbiased estimates of effect from the primary studies. Multiple testing and multiple modeling produce false, small p-values and biased estimates of treatment effects, which in turn make the meta-analysis unreliable/invalid.

## 2.0 Reliability of observational studies

### 2.1 Observational Studies

Epidemiology exhibits a notoriously poor record of reproducibility of published findings going back at least as far as Feinstein (4) and Mayes et al. (5); they found papers on both sides of the questions which were current at that time. There are continuing complaints about the reliability of observational studies: Taubes and Mann (6), Ioannidis (7), Kaplan et al. (8), Young and Karr (9), and Breslow (10, 11) to name a few. Breslow commented that "*contradictory results emanating from a plethora of irreproducible observational studies have contributed to the lack of esteem with which epidemiology is regarded by many in the wider biomedical community.*" Even the popular press is speaking up; Taubes (12) and Hughes (13) are two examples. See also Wikipedia (14) Replication crisis. Ominously, there may be actual misuse and/or even deliberate abuse of model fitting methods; see Glaeser (15), Young and Karr (9). Two groups of researchers using the same observational data base found that a treatment both caused, Cardwell et al. (16), and did not cause, Green et al. (17), cancer of the esophagus. Recently Schoenfeld and Ioannidis (18) found similar problems with nutritional studies, a review of ingredients in a cookbook.

The state of published scientific claims is sufficiently suspect that a consumer should start with the presumption that any claim made is as likely as not to be wrong (it will fail to replicate) and any treatment effect is likely biased. False positive p-values and biased estimates of treatment effects come in pairs.

2.2 The nature of a meta-analysis

In a meta-analysis, an effect estimate (statistic), e.g. a mean, a ratio, etc., is taken from each primary paper and combined, e.g. average, weighted average, to give a better measure of some finding, Chen and Peace (1) and Ehm (2). *Key requirements of the statistics (measures of effects) coming from primary papers used in a meta-analysis are that the statistics are independent and unbiased, Boos and Stefanski (3).* For a meta-analysis of randomized studies, these conditions are usually met; the papers are independent and randomization gives unbiased estimates. There still may be bias, e.g. non-adherence.

In randomized clinical trials (RCTs), very careful attention is given to the statistical analysis. A protocol, with a statistical analysis section (SAS) and a statistical analysis plan (SAP), is developed and agreed to by the interested parties, often a drug company and the FDA, *before* the study starts. One of the major concerns is the control of false positive results - a biased answer, which can be accomplished by using methods that preserve the overall experiment-wise error or false positive rate at a level of 5% or smaller.

Contrast the typical nutritional, Food Frequency Questionnaire, FFQ, observational study with a RCT. Each food is analyzed separately and sometimes in combinations. Multiple outcomes can be examined and the multiple food variables can be used as predictors. The analysis can be adjusted by putting multiple covariates into and out of the model, e.g. age, gender. There is usually no written protocol. For these factors (outcomes, predictors, covariates), there is no standard analysis strategy. The improvised strategy is often try-this-and-try-that. Under these circumstances, the analysis is essentially exploratory and the validity of claims/conclusions need further investigation.

**2.3 Our thesis**

Our thesis is that these FFQ papers are essentially exploratory - not confirmatory. In an exploratory study, typically many questions are at issue and analysis methods can be modified to meet the situation; in a confirmatory study, the entire study process is specified in a protocol, the outcome and predictor are specified, and statistical error is controlled, etc. In addition, there may be publication bias, negative studies are either not reported or they are rejected by editors and referees.

A major contribution of this research is to show that the analysis strategy used in the 10 primary papers produce biased statistics, which are unsuitable for inclusion in a meta-analysis study. Again, the potential for multiple testing and multiple modeling to bias results is well-known, but this problem is typically ignored in FFQ studies.

**3.0 Data**

Malik (19) with a meta-analysis looks at sugar-sweetened beverages and the risk of metabolic syndrome and type 2 diabetes, hereafter referred to as Malik. We examine the 10 primary papers (20-29) upon which the meta-analysis is based.

**4. Methods**

**4.1  Operation**

A protocol and data extraction form (DEF) were developed (Appendix) and the methods therein followed. Two teams were formed, each consisting of an Assistant Professor, a DrPH student and a Master's level student of Biostatistics.  Membership of the teams was determined randomly. The 10 primary papers were randomly assigned in a balanced fashion to the two teams.  Each team carefully reviewed the assigned 5 papers, extracted data therefrom in period 1 and filled in the DEF. The 5 papers reviewed by Team 1 were then crossed over to Team 2 for review and data extraction and vice versa. Differences in extraction results from the DEF between the two teams were resolved by the Co-PIs. A final DEF was completed for each paper. All final DEFs were posted to the study folder on the Google drive in PDF format. All final DEFs are available as supplemental material.

**4.2 Screening and Evaluation**

We read the meta-analysis and primary papers, filled in the data extraction form (DEF), and requested raw data from the lead author of each primary paper. Data extracted were: sample sizes, p-values, relative risks, confidence limits outcomes, predictors, covariates and funding sources. The counting of covariates is difficult as they can appear throughout the paper. Also, the number of potential covariates might be larger than the count as some are not mentioned in the paper.

Functions of these counts were used to estimate the potential size of the analysis search space available to the researcher; i.e. the number of comparisons and models available. The potential analysis search space for each primary paper was computed as the number of outcomes times the number of predictors times $2^c$, where c is the number of covariates.  This formula was applied to information found in both the abstract (Space A) as well as the text (Space T) of each primary paper.

**5.0 RESULTS**

Across the 10 studies (Table 1), sample sizes ranged from 4,304 to 91,249, with a median of 28,897 and a total of 332,357; smallest nominal P-Values ranged from 0.0001-0.001; and largest reported relative risk (RR) ranged from 1.23 – 5.06 with a median of 2.07. Eighty percent (80%) of the studies reported only government funding, 10% reported both government and non-government funding and 10% was unfunded.

The number of outcomes, predictors, and covariates for each of the 10 primary papers appear in Table 2. The range and median of the number of comparisons possible across the primary papers are (2 - 12,288) and 6.5, respectively for Space A (Table 3), and (3,072 - 117,117,952) and 196,608, respectively for Space T (Table 4). None of the 10 papers mention correcting for multiple testing or multiple modeling. There is no evidence of adjusting for multiplicities of any kind. All papers appear to test at the 5% level.

As it is impossible to "prove a negative," it is the responsibility of a researcher making a claim to provide strong evidence in support of the presumed positive claim. Given the multiple testing and multiple modeling, none of these papers provide strong evidence for their claims.

Any claim made could easily be a false positive. Note that each of the 10 primary papers should be examined separately for the validity of inferences. They must stand on their own before they can be considered for combining in a meta-analysis. As the statistics used in the primary papers do not provide valid evidence for the claim, the validity of the claim from the meta-analysis paper is questionable.

---

**Expected order statistics and corresponding p-values**

It is useful to review order statistics, their expected values and their relation to expected p-values as a function of the number of observations in a sample. If a random sample is taken from a population, and the objects are ordered from smallest to largest, the reordered objects are called order statistics. The value of the largest order statistic in the sample does not change from its value in the unordered sample, but it is different. It is the largest number in the sample. The larger the sample, the larger is the expected value of the largest object (**see Table 5**). Consider a sample from the normal distribution with a mean of zero and a standard deviation of one. If there are 10 objects in the sample, then the expected value of the largest object is 1.54.

Besides the expected value, the p-value for a z-test against the value zero is given. We expect the largest value in a sample of 10 to be about 0.5 standard deviations above the mean. Now look down Table 5. As sample size (N) increases, the expected value of the largest order statistic increases. In a sample of 30, we expect the largest order statistic, by chance alone, to be about 2.04 standard deviations larger than the mean (of zero). The corresponding (unadjusted) p-value is 0.0495, which would be nominally statistically significant.

As the numerical value of the largest order statistic is the same as its value in the sample, the fact that it is an order statistic is often overlooked, which we consider a grave mistake as it is biased; i.e. its expected value is not zero. It is wrong to consider the largest order statistic as if it were a random observation! It is common to adjust p-values when there are many questions at issue to control the false positive error rate. This Table can be used to remind a researcher that the value of an object can be large by chance alone and that a p-value can be small, again by chance alone. The larger the sample, the larger will be the value of the largest order statistic and the smaller will be the p-value.

If, after adjustment, a p-value is not statistically significant, the researcher needs to keep in mind that the corresponding experimental value is an order statistic and needs to be judged by expected values of order statistics, not as if it were a random value from the distribution in question.

The researcher can "cut" a continuous distribution to create ordered groups. The low group can be used as a reference group and the other groups can be compared to the reference group. The set of groups can be tested for linear trend. In Table 7, the number of p-values displayed in each paper is given as #Tests.

To make things concrete, suppose that there are 60 questions at issue in a study, and to simplify this discussion, assume the questions are independent of one another, then by chance alone we would expect to see a mean value of 2.31 standard deviations and a p-value of 0.027 for the largest order statistic. Without taking order statistics and multiple testing into account, we would declare statistical significance AND we would not expect the result to replicate. We would have a false positive. Taking the value of 2.31 to a meta-analysis would bias the meta-analysis and therefore would be misleading. The classic reference for computing the expected value of order statistics is Royston (30) who also quotes a well-known formula by Blom (31).

Statistics from Table 1 of Malik (19) were extracted and placed in our Tables 6 and 7. Using their Risk Ratios and Confidence Limits we computed z-tests, unadjusted and adjusted p-values. Note that in our Tables 6 and 7, we have two rows for Nettleton, one for diabetes and one for metabolic syndrome. After Bonferroni adjustment, there are no statistically significant results, which implies that the observed risk ratios are biased. With the raw data, we could use resampling to compute more accurate adjustments, Westfall and Young (32).

Table 7 gives more characteristics related to statistical testing. We note the number of foods used in the food frequency questionnaires, FFQs. The number varies from a low of 61 to a high of 165. Each of these foods could be used individually or in combination as a predictor of the health effect. The number of covariates given explicitly in Table 2 of Malik (19) is given as Total. Note that their counts of covariates are smaller than the number of covariates we counted (Table 4). Clearly, the number of covariates mentioned in the abstract (Table 4), is an underestimate of the number of covariates in play.

It is common to group the predictors. In this case, the number of groupings varies from 3 to 5. Using these groupings, the researchers tested for a linear trend, outcome versus ordered group, and they also tested the highest group against the lowest group so there were two dose response tests. The number of groupings could be selected to affect the results. Finally, each paper presents many reported p-values, #Tests. None of the primary papers reported any adjustment for multiple testing or multiple modeling, nor did they point to a registered protocol.

**DISCUSSION**

The first point to make is that the authors of the primary papers were, in effect, doing exploratory analyses. The analysis search space for each paper was vast and nominal statistical significance of 5% is, at best, a screen, not confirmatory in any sense. A major multiple testing dilemma occurred in the 1990s when genomics came on line. Lander and Kruglyak (32) argued that for claims to be believable there should be multiple testing correction over the entire analysis search space. None of the ten primary papers we examined performed any adjustment for multiple testing or multiple modeling and that appears to be the norm for analysis of Food Frequency Questionnaires, FFQs. By listing multiple outcomes, multiple predictors, and multiple covariates, the authors of these ten papers essentially declare their work exploratory. As the authors of the primary studies pointed to no protocol, provided no data, and made no multiplicity or modeling adjustments, the reader is left without much recourse other than to consider the papers exploratory.

Two recent books by Harris (34) and Chandler (35) note that multiple testing, which they term p-hacking, and justifying a test of hypothesis AFTER looking at the data, HARKing (Hypothesizing After the Results are Known) are very common and of course upset the validity of p-values and would bias any associated statistics.

Here is a missing insight. <u>In real science, a hypothesis is refined and then retested with new data on a more restricted question.</u> The protocol is written before the new data is analyzed. There is statistical error control. There is often replication. We should give greater credence to the results of the new, more definitive study. If it is positive, we say the hypothesis is supported, standard Karl Popper reasoning. If the new study fails, we consider abandoning the hypothesis and spend science resources on some other problem.

In FFQ studies there are ~60 to ~130 food questions and many of these food questions are repeated from one study to another. If the covariates are fixed in a protocol so they do not introduce model variation, then nutrition studies that use FFQs in addition to making claims for positive findings, offer an opportunity for replicating findings (or not) from other studies. The statistical analysis of each food is easily accomplished with a few lines of code. A p-value plot would facilitate examination of all the questions, Schweder and Spjøtvoll (36) as was done in Young et al.(37).

It is rather routine for a researcher not to submit negative papers for publication as the belief is that editors are likely to reject negative papers. Informal conversations with multiple authors of published negative studies support the difficulty of getting them published. Across the board, negative studies have a more difficult time getting published. Given that negative papers are typically not published, eventually, we can have serious publication bias, positive studies are accepted as they support the current paradigm and negative studies are rejected. As far as we know, observational studies used as primary studies in meta-analyses are not routinely examined for multiple testing and multiple modeling biases. For more discussion of publication bias see Wikipedia (14), Publication bias.

Humans like good stories, which becomes a useful art in the writing of a scientific paper. Authors can accentuate positive papers and downplay or even omit negative papers, Kabat (38). It is very easy for presumptively neutral researchers to become believers in their own claim, Feynman (38) among many others, or an existing popular paradigm, e.g. Kuhn (39).

Kuhn (39) noted that it is very difficult to overturn an existing scientific paradigm. Those doing nutrition and health effects research should be held to strict scientific standards: state if a study is exploratory, refine claims coming from an exploratory study for a confirmatory study, and most importantly, make data sets and analysis code available, etc.

Scientifically and logically, it is not possible to prove a negative so to make a public health claim, an investigator should provide strong evidence - an analysis that names all the questions at issue, and fairly adjust for multiple testing and multiple modeling. None of the claims made in the 10 primary papers can be considered reliable due to potential bias, and hence they are inappropriate for inclusion in a meta-analysis.

We, the science community, are not recognizing that authors are doing exploratory data analysis over and over, year after year. They look at multiple outcomes, multiple causes, any number of covariates, and any number of predictors. They try this and try that analysis and publish a paper if they get a p-value less than 0.05 where a plausible story can be made (34, 35, 38). If they fail to find "statistical significance," then it appears that they simply do not publish. Those doing meta-analyses need to realize the problem to their work. Authors, editors, and consumers can become true believers in a false paradigm.

Finally, the lead author of each of the 10 primary papers was contacted twice asking if data used in their paper were available. None of the authors provided their analysis data set. Unfortunately, it is common for authors not to provide their analysis data set. Without access to the data sets, it is not possible to adjust the analysis for multiple testing and multiple modeling. From what is available in the papers and as summarized in Table 1 of Malik (19), it appears that none of the claims made in the 10 primary papers would be statistically significant after adjustment. The data should be made public so that the analyses can be corrected for the bias introduced by multiple testing and multiple modeling.

**SUMMARY**

Ten primary papers used in the meta-analysis study by Malik et al. (19) were carefully examined with respect to the range of analysis options open to the researcher, and the size of the analysis search space. The search space for each paper is large (in many cases vast) considering all the questions possible, so that testing claims at a nominal 0.05 is problematic. Meta-analysis using these papers should also be considered unreliable until the reliability of the underlying primary papers is assessed or confirmatory studies are run.

**ACKNOWLEDGMENTS:**

Karl Peace substantially contributed to design of the study, analysis of the data, writing the manuscript as well as major editing of the manuscript.

JingJing Yin contributed to design of the study, analysis of the data, writing the manuscript as well as major editing of the manuscript.

Haresh Rochani contributed to the design of the study, analysis of the data, writing the manuscript as well as major editing of the manuscript.

Sarbesh Pandeya contributed to the design of the study and analysis of the data.

Stanley Young contributed to design of the study, analysis of the data, writing the manuscript as well as major editing of the manuscript.

We also would like to thank our students, Kotwoallama Reine Zerbo, Patrick Chang and Yi Hao who helped in data extraction process for this paper.

**Table 1:** Review of Sample Size for 10 Primary papers

| Paper ID | Overall Sample Size | Sample Size per Group |
|---|---|---|
| Nettleton et al. (2009) | 6,814 | Rare or never: 2961, >rare/never but < 1 servings per week: 455, >= 1 servings/week to < 1 servings/day: 914, >=1 serving/day: 681 |
| Lutsey et al. (2008) | 9,514 | Men: 4197, Women: 5317 |
| Dhingra et al. (2007) | 8,997 | <1 soft drinks per day: 5840, 1 soft drinks per day: 1918, >= 2 soft drinks per day: 1239 |
| Montonen et al. (2007) | 4,304 | 1st quartiles: 1076, 2nd quartile: 1076, 3rd quartile: 1076, 4th quartile: 1076 |
| Paynter et al. (2006) | 12,204 | Men: 5414, Women: 6790 |
| Schulze et al. (2004) | 91,249 | For 1991, <1/mo: 49203, 1-4/mo: 23398, 2-6/wk: 9950, <1/d: 8698; For 1991-1995, <=1/wk: 38737, >=1/d: 2366, <=1/wk to >=1/d: 1007, >=1/d tp <=1/wk: 1020 |
| Palmer et al. (2008) | 43,960 | Soft drinks per week: <1: 25971, 2-6: 10521, >=1: 7468; Fruit Drinks per Week : <1: 15455, 2-6: 13722, >=1: 13644 |
| Bazzano et al. (2008) | 71,346 | quintile 1: 14573, quintile 2: 14408, , quintile 3: 14337, , quintile 4: 14118, , quintile 5: 13913 |
| Odegaard et al. (2010) | 43,580 | Soft drink consumption: almost never: 32060, 1-3/Month: 4514, 1/week: 2389, 2-3/week: 4617; Juice Consumption: almost never: 35719, 1-3/Month: 4399, 1/week: 1791, 2-3/week: 1671 |
| De Koning et al. (2010) | 40,389 | Sugar Sweetened beverages: Q1: 13675, Q2: 5022, Q3: 11729, Q4: 9963; Artificially sweetened beverages: Q1: 18442, Q2: 2681, Q3: 9448, Q4: 9818 |
| **Across all articles** | **332,357** | **Number of Groups: 2,2,3,(3,3),4,(4,4),(4,4),(4,4),5** |

**Table 2:** P-values, Relative Risks, Multiplicity Adjustment & Funding Source for 10 Primary papers

| Paper ID | Smallest p-value | Largest RR (Hazard ratio) | Largest RR: CI | Multiplicity Adjustment for p-values | Funding Source |
|---|---|---|---|---|---|
| Nettleton et al. (2009) | <0.001 | 2.2 | (1.1-4.51) | No | Government |
| Lutsey et al. (2008) | <0.001 | 1.34 | (1.24-1.44) | No | Government |
| Dhingra et al. (2007) | <0.0001 | 2.31 | (1.77-3.01) | No | Government and Non-government |
| Montonen et al. (2007) | <0.001 | 5.06 | (1.87-3.71) | No | Unfunded |
| Paynter et al. (2006) | <0.01 | 1.23 | (0.93-1.62) | No | Government |
| Schulze et al. (2004) | <0.001 | 2.31 | (1.55-3.45) | No | Government |
| Palmer et al. (2008) | 0.001 | 1.51 | (1.31-1.75) | No | Government |
| Bazzano et al. (2008) | <0.001 | 4.47 | (2.35-7.66) | No | Government |
| Odegaard et al. (2010) | <0.0001 | 1.7 | (1.34-2.16) | No | Government |
| De Koning et al. (2010) | <0.01 | 1.94 | (1.75-2.14) | No | Government |
| **Across all articles** | **<0.0001 - <0.01** | **1.23 - 5.06** | | | **90% Government** |

**Table 3:** Search space Size of 10 Primary papers based on **Abstracts**

| Primary papers | Primary papers Journals | Outcomes | Predictors | Covariates | Space Size |
|---|---|---|---|---|---|
| Nettleton et al. | Diabetes Care 2009 | 2 | 1 | 3 | 24 |
| Lutsey et al. | Circulation 2008 | 1 | 2 | 4 | 32 |
| Dhingra et al. | Circulation 2007 | 7 | 1 | 10 | 7,168 |
| Montonen et al. | J Nutr 2007 | 1 | 5 | 0 | 5 |
| Paynter et al. | Am J Epidem 2006 | 1 | 2 | 0 | 2 |
| Schulze et al. | JAMA 2004 | 2 | 1 | 2 | 8 |
| Palmer et al. | Arch Intern Med 2008 | 1 | 1 | 2 | 4 |
| Bazzano et al. | Diabetes Care 2008 | 1 | 3 | 0 | 3 |
| Odegaard et al. | Am J Epidem 2010 | 1 | 2 | 2 | 8 |
| de Koning | Am J Epidem 2011 | 1 | 3 | 10 | 12,288 |

5**Table 4:** Space Size of 10 Primary papers based on **Texts of Papers**

| Primary papers | Primary papers Journals | Outcomes | Predictors | Covariates | Space Size |
|---|---|---|---|---|---|
| Nettleton et al. | Diabetes Care 2009 | 2 | 2 | 15 | 196,608 |
| Lutsey et al. | Circulation 2008 | 1 | 2 | 14 | 32,678 |
| Dhingra et al. | Circulation 2007 | 7 | 1 | 24 | 117,117,952 |
| Montonen et al. | J Nutr 2007 | 1 | 12 | 15 | 392,396 |
| Paynter et al. | Am J Epidem 2006 | 1 | 2 | 14 | 32,678 |
| Schulze et al. | JAMA 2004 | 2 | 3 | 9 (Mod 1) | 3,072 |
| Palmer et al. | Arch Intern Med 2008 | 2 | 3 | 15 | 196,608 |
| Bazzano et al. | Diabetes Care 2008 | 1 | 5 | 13 | 40,960 |
| Odegaard et al. | Am J Epidem 2010 | 2 | 2 | 16 | 262,144 |
| de Koning | Am J Epidem 2011 | 1 | 3 | 24 | 6,291,456 |

**Table 5:** Sample size, Expected value of largest Order Statistics and corresponding P-Values. Note that the value of the largest order statistics for N of 1000 and 5000 was computed using a formula of Blom (1958).

| N | Exp. Value of largest Order Statistic | P-Value |
|---|---|---|
| 10 | 1.53875 | 0.12211 |
| 20 | 1.86748 | 0.06976 |
| 30 | 2.04276 | 0.04952 |
| 40 | 2.16078 | 0.03864 |
| 50 | 2.24907 | 0.03181 |
| 60 | 2.31928 | 0.02709 |
| 70 | 2.37736 | 0.02364 |
| 80 | 2.42677 | 0.02099 |
| 90 | 2.46970 | 0.01890 |
| 100 | 2.50759 | 0.01720 |
| 125 | 2.58634 | 0.01407 |
| 150 | 2.64925 | 0.01194 |
| 175 | 2.70148 | 0.01038 |
| 200 | 2.74604 | 0.00919 |
| 225 | 2.78485 | 0.00826 |
| 250 | 2.81918 | 0.00750 |
| 300 | 2.87777 | 0.00635 |
| 350 | 2.92651 | 0.00551 |
| 400 | 2.96818 | 0.00487 |
| 1000 | 3.24144 | 0.00119 |
| 5000 | 3.67755 | 0.00024 |

**Table 6**: Risk Ratios, Confidence Limits taken from Table 1 of Malik et al. 2010. Z-tests, p-values, adjustment factors and adjusted p-values were computed.

| Ref | Sig | RR | CLL | CLH | Beta | BetaSE | Z | Prob | AdjFactor | AdjP |
|---|---|---|---|---|---|---|---|---|---|---|
| Nettleton et al. | 0.05 | 0.86 | 0.62 | 1.17 | -0.151 | 0.162 | -0.931 | 0.8241 | 116736 | 1.000 |
| Lutsey et al. | p-val | 1.09 | 0.99 | 1.19 | 0.086 | 0.047 | 1.836 | 0.0332 | 540672 | 1.000 |
| Dhingra et al. | CL 95% | 1.39 | 1.21 | 1.59 | 0.329 | 0.070 | 4.726 | <0.0001 | 244 | 0.000 |
| Montonen et al. | CL 95% | 1.67 | 0.98 | 2.87 | 0.513 | 0.274 | 1.871 | 0.0307 | 102400 | 1.000 |
| Paynter et al. | CL 95% | 1.17 | 0.92 | 1.39 | 0.122 | 0.157 | 1.491 | 0.0679 | 244 | 1.000 |
| Schulze et al. | 0.05 | 1.83 | 1.42 | 2.36 | 0.604 | 0.130 | 4.663 | <0.0001 | 2179072 | 1.000 |
| Palmer et al. | CL 95% | 1.24 | 1.06 | 1.45 | 0.215 | 0.080 | 2.692 | 0.0036 | 2228224 | 1.000 |
| Bazzano et al. | CL 95% | 1.31 | 0.99 | 1.74 | 0.270 | 0.144 | 1.877 | 0.0303 | 11264 | 1.000 |
| Odegaard et al. | CL 95% | 1.42 | 1.25 | 1.62 | 0.351 | 0.066 | 5.301 | <0.0001 | 1351680 | 0.078 |
| de Koning | 0.05 | 1.14 | 1.03 | 1.28 | 0.131 | 0.055 | 2.364 | 0.0090 | 8384 | 1.000 |
| Nettleton et al.* | 0.05 | 1.15 | 0.92 | 1.42 | 0.140 | 0.111 | 1.262 | 0.1034 | 116736 | 1.000 |

*This was not included in the pool of ten primary papers that but was reported by Malik. Et al (2010).

**Table 7:** Number of foods, FFQ, considered in each primary paper, the Total number of covariates. TCovars, the number of groupings used for predictors and the type of statistical testing – based on Table 1 of Malik. Also given are the number of p-values reported, #Tests, derived by counting in each primary paper.

| Ref | FFQ | TCovars | #groups | Method | #Tests |
|---|---|---|---|---|---|
| Nettleton * | 114 | 10 | 4 | Trend, Each vs control | 88 |
| Lutsey | 66 | 13 | 5 | Trend, Each vs control | 85 |
| Dhingra | 61 | 2 | 3 | Trend, Each vs control | 101 |
| Montonen | 100 | 10 | 4 | Trend, Each vs control | 63 |
| Paynter | 61 | 2 | 5 | Trend, Each vs control | 60 |
| Schulze | 133 | 14 | 4 | Trend, Each vs control | 54 |
| Palmer | 68 | 15 | 3 | Trend, Each vs control | 87 |
| Bazzano | 88 | 7 | 5 | Trend, Each vs control | 114 |
| Odegaard | 165 | 13 | 4 | Trend, Each vs control | 50 |
| de Koning | 131 | 6 | 4 | Trend, Each vs control | 84 |

*This was not included in the pool of ten primary papers but was reported by Malik, et al (19). His references are given as our references (20-29).

**Appendix: Protocol V02: MMA Study**

**Note:** Sections of Protocol V01 were rewritten upon discovering that the Malik et al paper had only 10 non-overlapping primary papers.

**Co-PI:** Karl Peace, Jiann-Ping Hsu College of Public Health, Georgia Southern University, kepeace@georgiasouthern.edu

**Co-PI:** Stan Young, CGStat, genetree@bellsouth.net

**Background:** For many nutritional questions randomized trials are not available so observational studies are conducted. It is common to gather a number of observational studies related to a question. The individual studies are evaluated and summary results from the studies are combined using what is called meta-analysis methods.

**Idea:** Our study is to evaluate the reliability of a nutritional meta-analysis study by examining the statistical reliability of the underlying studies.

The meta-analysis study of Malik, et al. (19) was selected for study. Within the paper there appeared to be 11 cited base studies. However, upon examination of Dr. Young, one appeared a replicate. Hence only the 10 non-overlapping primary papers were reviewed and contained data for extraction.

**Objectives:**

a. Determine the size of the analysis search space for each observational base study of a meta-analysis.
b. Determine if uncorrected summary statistics invalidate meta-analysis claims.

**Study Population:** Primary papers from a meta-analysis paper of observational studies.

**Locating studies:** Reference list from the meta-analysis paper

**Screening and Evaluation Methods:**

a. Read meta-analysis and primary papers.
b. Fill in Data Extraction Form
**c.** Ask for data access.

**Operation:**

Two teams will be formed, each consisting of an Assistant Professor of Biostatistics, a DrPH student and a Master's level student. Membership of the teams will be determined randomly.

The 10 primary papers will be randomly assigned in balanced fashion to the two teams. Each team will review and extract data from the assigned 5 papers during period 1. The 5 papers reviewed by Team 1 will then be crossed over to Team 2 for review and data extraction and vice versa. Differences in extraction results between the two teams will be resolved by the Co-PIs. A final Data Extraction Form will be completed for each paper. All Data Extraction Forms will be posted to the study folder on the Google drive in PDF format.

The search space will be computed for each primary paper as:

#outcomes x #predictors x $2^c$, where c is the number of covariates in the final model.

**Results:** The summary results for a paper will be considered unreliable if the search space is greater than 100 or if #outcomes x #predictors is greater than 10. The meta-analysis paper will be considered unreliable if over ¼ if the primary papers are considered unreliable.

**References:**

To minimize print space, the Malik et al paper is **reference 19** and the 10 primary papers are **references 20-29** of the References section of the manuscript.

**Data Extraction Form      Preliminary    Final**

**MMA Study: Malik et al. Diabetes Care 2010; 33, 2477-2483.**

Your name_______________________     Date________________

1. Paper (fill in the literature references as it appears in the meta-analysis paper)

2. PI: name, email address, regular mail

3. Journal editor: name, email address

4A. Overall Sample size: _____________

4.B  Sample size per Group (identify group)

    Group 1: ______________________     Sample Size___________

    Group 2: ______________________     Sample Size___________

    Group 3: ______________________     Sample Size___________

    Group 4: ______________________     Sample Size___________

5. Smallest p-value _________     Largest RR with CL ________________

6. # outcomes      From Abstract_____     From Paper_____

7. # predictors      From Abstract_____     From Paper_____

8. # covariates      From Abstract_____     From Paper_____

8.A. #  potential covariates mentioned _____________

8.B. # Covariates used in the analysis model _____________

9. Is a food questionnaire used in the study?  Yes   No

10. Raw Data available (as stated in the paper)?  Yes   No

11. Funding source.  Government Grant Number________     Industry______     Unfunded

12. Eligibility Criteria

13. Comments. Any other things of potential interest noted while reviewing the paper.